\documentclass{ws-procs9x6}

\begin{document}

\title{NEW CLASS OF QUANTUM DEFORMATIONS OF $D=4$ EUCLIDEAN SUPERSYMMETRIES
\footnote{Proceedings of the XXIX International Colloquium on Group-Theoretical Methods in Physics, Tianjin August2012;
Ed.Chengming Bai, Jean-Pierre Gazeau, Mo-Lin Ge, World Scientific, Singapore 2013.}}

\author{A. BOROWIEC, J. LUKIERSKI$^{**}$ and M. MOZRZYMAS}

\address{Institute for Theoretical Physics, Wroclaw University,\\
pl. Maxa Borna 9, 50-204 Wroclaw, Poland\\
$^{**}$E-mail: lukier@ift.uni.wroc.pl\\
www.ift.uni.wroc.pl}

\author{V. N. TOLSTOY}

\address{Lomonosov Moscow State University, \\
Skobeltsyn Institute of Nuclear Physics\\
Moscow 119991, Russian Federation\\
E-mail: tolstoy@nucl-th.sinp.msu.ru}

\begin{abstract}
 We present the class of deformations of simple Euclidean superalgebra, which
describe the supersymmetrization of some Lie algebraic noncommutativity of $%
D=4$ Euclidean space-time. The presented deformations are generated by the
supertwists. We provide new explicit formulae for a chosen twisted $D=4$
Euclidean Hopf superalgebra and describe the corresponding quantum covariant
deformation of chiral Euclidean superspace.
\end{abstract}


\bodymatter

\section{Introduction}\label{sec1}
 In the applications of quantum deformations to fundamental interaction
theories it should be taken into considerations that supersymmetry is one of
the main basic building blocks of the string and M-theory. We recall that
the classification of deformations of $D=4$ Poincar\'{e} symmetries is almost
complete \cite{c1,c2} (see also Ref. \refcite{c3,c4}), but systematic studies of
supersymmetric extensions of deformed Poincare algebras quite recently were considered
\cite{c5,c6,c7} . In this note we shall deal with the superextensions of quantum $D=4$ Euclidean
symmetries providing the Lie-algebraic structure of corresponding Euclidean
space-time commutators. We add that Euclidean theories are often very
useful as intermediate step in the calculations in Minkowski space.

Historically first and the most known example of Lie-algebraic space-time
deformations are the $\kappa -$deformations of Minkowski space-time \cite{c8,c9,c10}%
 \begin{equation}\label{a1}
 \lbrack \widehat{x}_{\mu },\widehat{x}_{\nu }]=\frac{i}{\kappa }(a_{\mu }%
\widehat{x}_{\nu }-a_{\nu }\widehat{x}_{\mu })
\end{equation}
where $a_{\mu }=(1,0,0,0)$ corresponds to the standard (time-like) $\kappa $%
-deformation \cite{c11} and if $a_{\mu }=(1,1,0,0)$ we get the light-cone $\kappa $%
-deformation \cite{c12}.

In this  short paper we would like to consider an example of $D=4$
Euclidean supersymmetries, which in its bosonic sector provides the
Lie-algebraic deformation of $D=4$ space-time sector. This choice of
deformation follows from our recent paper \cite{c7} where we did present the
Euclidean counterparts of Zakrzewski list of $D=4$ Poincare classical $r$%
-matrices \cite{c2}  and classified their supersymmetrizations in the
case of simple chiral (holomorphic) Euclidean superalgebras \footnote{Simple
chiral (holomorphic) $D=4$ Euclidean superalgebra with four complex supercharges is
usually denoted as $N=(\frac{1}{2},\frac{1}{2})$ Euclidean SUSY; by Hermitean conjugation of supercharges
one obtains simple antichiral (antiholomorphic) $N=(\frac{1}{2},\frac{1}{2})$ superalgebra.
The lowest dimensional real Euclidean $N=(1,1)$ superalgebra is described by eight
complex supercharges satisfying $SU(2)$-Majorana reality conditions and it contains $N=(\frac{1}{2},\frac{1}{2})$
chiral and antichiral sectors.}.

Plan of our presentation is the following. In section 2 we deliver the list
of Euclidean $D=4$ classical $r$-matrices and their supersymmetrizations
presented in Ref. \refcite{c7}. We show that the conjugation and pseudoconjugation of Euclidean superalgebras
are related with quaternionic structure of fundamental $O(3)$ and $O(4)$ spinors.
In section 3 we give explicit formulae for particular $D=4$
Euclidean quantum twisted superalgebra, which leads to the Lie-algebraic deformation
of space-time sector. Further we provide the deformation of corresponding
Euclidean chiral $N=(\frac{1}{2},0)$ superspace and indicate the link with
Seiberg $N=\frac{1}{2}$ deformation of superspace. In section 4 we present
short outlook.
\section{$D=4$ Euclidean twisted deformations and their $N=\frac{1}{2}$
superextenstions}\label{ab:sec2}
\subsection{Euclidean twisted deformations}
 Let us recall $D=4$ Euclidean Lie algebra $E(4)=o(4)\ltimes P_{4}$, where
six Euclidean rotation generators $M_{\mu \nu }=-M_{\nu \mu }\in o(4)$ $(\mu ,\nu
=0,1,2,3)$ and Euclidean four-momenta $P_{\mu }\in P_{4}$ satisfy the
relations%
\begin{eqnarray}
\lbrack M_{\mu \nu },M_{\lambda \rho }] &=& i(\delta_{\nu \lambda}M_{\rho \mu }-\delta _{\nu  \rho}M_{\lambda \mu }
-\delta _{\mu \lambda }M_{\rho \nu }+\delta _{\mu \rho }M_{\lambda\nu }) \label{a2}\\
\lbrack M_{\mu \nu },P_{\rho }] &=&i(\delta_{\rho  \mu }P_\nu -\delta _{\rho \nu }P_\mu ),\qquad
\lbrack P_{\mu },P_{\nu }]=0. \label{a3}
\end{eqnarray}
The generators $(M_{\mu \nu },P_{\mu })$ are Hermitean. We recall that from
relations (\ref{a2})-(\ref{a3})  we can obtained  real $D=4$ Poincar\'{e} algebra $P(3,1)=o(3,1)\ltimes
\widetilde{P}_{4}$ ($(\widetilde M_{\mu \nu }\in o(3,1),$ $\widetilde{P}_{\mu }\in P_{4}$
where $\mu ,\nu =0,1,2,3)$ after the substitution \ $(i,j=1,2,3)$
\begin{equation}\label{a4}
\widetilde M_{ij}=M_{ij},\qquad \widetilde M_{i0}=-iM_{i0},\qquad \widetilde{P}_{i}=P_{i},\qquad
\widetilde{P}_{0}=-iP_{0}.
\end{equation}
Strictly speaking it means that one can introduce, on a complex algebra generated by the relations (\ref{a2})-(\ref{a3}),  two non-isomorphic real structures (i.e. $\star-$conjugation operations): the Euclidean one with respect to which the Euclidean generators $(M_{\mu\nu}, P_\rho)$ are Hermitean (i.e. selfconjugated) and the Poincar\'{e} one with $(\widetilde M_{\mu\nu}, \widetilde P_\rho)$ Hermitean. Both conjugations (real forms) coincide on the generators $(M_{i j}, P_k)$ \cite{c7}.

In Ref. \refcite{c7} we observed that all the real Poincar\'{e} $r$-matrices introduced by S.
Zakrzewski in Ref. \refcite{c2} describe equally well the classical $r$%
-matrices for complex $D=4$ Euclidean Lie algebra $o(4;\mathbb{C})\ltimes P_{4}^{\mathbb{C}}$
which describes also $D=4$ Poincar\'{e} algebra.
It appears that out of the Zakrzewski list with 21
classical $r$-matrices only 8 admits the reality conditions leading to $D=4$
Euclidean real classical $r$-matrices. The list of such matrices is the
following (we use $M_{i}=\frac{1}{2}\epsilon _{ijk}M_{jk},$ \ $N_{i}=M_{0i},
$ \ $P_{\pm }=P_{0}\pm P_{3})$
\begin{eqnarray}
r_{1}&=&i\xi N_{3}\wedge M_{3}+\zeta_1 P_{+}\wedge P_{-}+\zeta_2%
P_{1}\wedge P_{2}\nonumber\\
r_{2}&=&\xi P_{0}\wedge M_{3}+r_{8},\qquad r_{3}=\xi P_{3}\wedge
N_{3}+r_{8},\qquad r_{4}=\xi P_{1}\wedge M_{3}+r_{8},\nonumber\\
r_{5}&=&\xi P_{+}\wedge M_{3}+r_{8},\qquad r_{6}=\zeta
_{1}P_{1}\wedge P_{+},\label{a5}\\
r_{7}&=&\zeta _{2}P_{1}\wedge P_{2},\qquad r_{8}=\zeta _{1}P_{0}\wedge
P_{3}+\zeta _{2}P_{1}\wedge P_{2},\nonumber
\end{eqnarray}
where the parameters  $\xi ,$  $\zeta _{1},$ $\zeta _{2}$ are real.

It is interesting to deduce from (\ref{a5}) that the classical $r$-matrix
describing standard (Poincar\'{e})
light-cone $\kappa-$deformation does not have its Euclidean counterparts.

\subsection{Quaternionic structure of Euclidean spinors, Euclidean
supercharges conjugation and pseudoconjugation.}

The supercharges of Euclidean supersymmetry are described by fundamental $O(4)$ spinors, which are $O(4)$ Clifford algebra modules.
 Because $o(4)=o(3)\oplus o(3)$ let us consider firstly $o(3)$ spinors. Following the list of Clifford modules and fundamental
 spinors for orthogonal groups \cite{c13,RC,c14,JL} they are
described by single real quaternion $q=q_{0}+q_{r}e_{r}$, where $e_{r}$ describe
the quaternionic units $(e_{r}e_{s}=-\delta _{rs}+\epsilon _{rst}e_{t})$ and
$q_{0},$ $q_{r}$ are real. The irreducible fundamental representation
of spinorial covering $\overline{SO(3)}=Sp(1)\equiv U(1;H)=SU(2)$ is
described by real unit quaternions $q\overline{q}=q_{0}^{2}+q_{r}^{2}=1,$ $(%
\overline{q}=q_{0}-q_{r}e_{r})$. We recall that in physics one uses rather complex realization of quaternions.
In order to define a complex realization $%
\mathbf{q}$ of real quaternions we should firstly introduce matrix realizations $%
e_{r}=-i\sigma _{r}$, where $\sigma _{r}$ are $2\times 2$ Pauli matrices and
one obtains unit quaternions as decribing $2\times 2$ $SU(2)$ matrices
\begin{equation}\label{a6}
U= \left(
\begin{array}{cc}
q_{0}-iq_{3} & -q_{2}-iq_{1} \\
q_{2}-iq_{1} & q_{0}+iq_{3}%
\end{array}%
\right) =\left(
\begin{array}{cc}
z_{1} & -z_{2}^{\ast } \\
z_{2} & z_{1}^{\ast }%
\end{array}%
\right)
\end{equation}
where $z_{1}=q_{0}-iq_{3},$ $z_{2}=q_{2}-iq_{1}$ and $\det U\equiv |z_1|^2+|z_2|^2=1$. Here $z^\ast$ denotes
a complex conjugation of $z\in\mathbb{C}$.
The  one-component quaternionic spinor $q$ can be represented as two-component complex spinor, described by
the first column of the relation (\ref{a6} )
\begin{equation}\label{a7}
\mathbf{q}\mathit{=}
\left(
\begin{array}{c}
q_{0}-iq_{3}   \\
q_{2}-iq_{1} %
\end{array}%
\right) =\left(
\begin{array}{c}
z_{1}  \\
z_{2}  %
\end{array}  %
\right)\in \mathbb{C}^2
\end{equation}
The quaternionic nature of complex spinors (\ref{a7}) is exhibited by the presence of quaternionic pseudoconjugation ($\epsilon\equiv i\sigma_2$) \cite{c14,JL}
\begin{equation}\label{a7b}
 (\mathbf{q}^H)_\alpha =  \epsilon_{\alpha\beta}\mathbf{q}^\ast_\beta\qquad  (\mathbf{q}^{H})^H =-\mathbf{q}
\end{equation}
which commutes with $SU(2)\simeq SO(3)$ rotations, i.e. $(U\mathbf{q})^H=U\mathbf{q}^H$.
The condition (\ref{a7b}) is a complex realization of the quaternionic map
$q\rightarrow q^H=-qe_2$ \footnote{This quaternionic map as well as  its complex counterpart given by (\ref{a7b})
is not unique. One can replace $e_2\rightarrow \alpha_ie_i$ (with $\Sigma |\alpha_i|^2=1$)  in $q^H=-q e_2$.},
because the group $U(1;H)=SU(2)$ acts on $q$ from the left, its commutativity with the mapping $q\rightarrow q^H$
is obvious. In order to interpret the relation (\ref{a7b}) in a quaternionic framework one can deduce from (\ref{a7}) that
complex conjugation $\mathbf{q}\rightarrow\mathbf{q}^\ast$ represents the quaternionic map $q\rightarrow -e_2qe_2$,
and one gets $q^H=-qe_2= -e_2(-e_2qe_2)$, where in complex realization (\ref{a7b}) we inserted $e_2=-i\sigma_2$.

The Euclidean supercharges for general $D=4$ Euclidean $(n,m)$ superalgebra $%
(n,m$ are half-integers) are described by $n$ left  chiral $SU_L(2)$
spinors denoted  by $Q_{L\alpha}^{a}$ $(a=1,...,n,$ $%
\alpha =1,2)$ and $m$ right  chiral $SU_R(2)$ spinors denoted  by
  $Q_{R\alpha }^{b}$ $(b=1,..,m)$. In order to obtain the simplest $%
D=4$ Euclidean chiral (holomorphic) $N=(\frac{1}{2},\frac{1}{2})$ superalgebra we
need a doublet of complex supercharges $(Q_{L\alpha},Q_{R\beta })$ which
enters into the basic SUSY relation (see e.g. Ref. \refcite{c15} where is denoted $Q_{L\alpha}\equiv Q_{\alpha;}, Q_{R\alpha}\equiv Q_{;\alpha}$)%
\begin{eqnarray}
\{Q_{L\alpha},Q_{R\beta}\}&=&2(\sigma ^E_{\mu })_{\alpha\beta}\mathcal{P}^{\mu } \label{a8}\\
\{Q_{L\alpha ;},Q_{L\beta}\}&=&\{Q_{R\alpha },Q_{R\beta }\}=0 \label{a9}
\end{eqnarray}
where $Q_{L\alpha}, Q_{R\beta}$ transforms respectively under $SU_L(2), SU_R(2)$ and $\mathcal{P}^{\mu }$ are complexified fourmomenta. Further
\begin{equation}\label{b2}
    (\sigma_\mu^E) = (iI_2, \sigma_i)
\end{equation}
describe $D=4$ Euclidean Pauli matrices which transform under $D=4$ Euclidean rotations $\widetilde{SO(4)}=SU_L(2)\otimes SU_R(2)$
as Euclidean fourvector ($U_{L,R}\in SU_{L,R}(2)$)
\begin{equation}\label{b3}
    U_L\sigma^E_\mu U_R^T=A_\mu^\nu\sigma^E_\nu
\end{equation}
where real $4\times4$ matrices $A_\mu^\nu$ represent $SO(4)$ four-vector rotations. This is due to the fact that the Euclidean Pauli matrices (\ref{b2}) satisfy the constraints
 \begin{equation}\label{d2}
    (\sigma^E_\mu)^\ast=\epsilon\,\sigma^E_\mu\, \epsilon
\end{equation}
while for $SU(2)$ matrices (\ref{a6}) one has $U^\ast=-\epsilon\, U\epsilon$ instead.
This can be used to embed Euclidean space-time onto quaternions.
 We point out that the superalgebra (\ref{a8}-\ref{a9}) cannot be constrained by any reality condition, because
the supercharges $Q_{L\alpha}$ ,$Q_{R\beta }$ transform under two independent $SU(2)$ groups.

From physical point of view it would be desirable to obtain in the superalgebra (\ref{a8}) the real (Hermitean) fourmomenta $\mathcal{P}_\mu$. For that purpose one can adopt the quaternionic pseudoconjugation (\ref{a7b}) to the Euclidean supercharges
in the following two ways:

i) One introduces an abstract inner $SU(2)-$invariant  graded antiautomorphism of fourth order ($\tau^2=-1\Rightarrow\tau^4=1$) \footnote{It is well known that it does not exist $SU(2)-$invariant antiautomorphism of second order (conjugation) in two dimensions, i.e. matrices (\ref{a6}) cannot be made real at the same time by change of basis in $\mathbb{C}^2$.}
\begin{equation}\label{d1}
    \tau(Q_{L\alpha})=\epsilon_{\alpha\beta}Q_{L\beta}\qquad  \tau(Q_{R\alpha})=\epsilon_{\alpha\beta}Q_{R\beta}
\end{equation}
which  satisfies   the antilinearity property: $\tau(z Q)=\bar z\tau(Q)$ and extends the Euclidean Hermitian conjugation from the bosonic sector into the fermionic. The pseudoconiugation (\ref{d1}) was introduced by Nahm, Rittenberg, Scheinert \cite{c16}, Berezin, Tolstoy \cite{c17} and Manin \cite{c18} (see also \cite{c19,c20}).  Now one can check explicitly (taking into account (\ref{d1}) as well as the graded antihomomorphism property: $\tau(Q_1Q_2)=-\tau(Q_2)\tau(Q_1$)) that the relation (\ref{a8}-\ref{a9}) remains  invariant under pseudoconjugation (\ref{d1}) if the Euclidean fourmomentum generators are real.

ii) one can assume that the supercharges $Q_{L\alpha}, Q_{R\beta}$ admit well defined Hermitean conjugation which is an antilinear involutive outer map provided linearly independent conjugated antichiral generators $Q^+_{L\alpha}\equiv Q_{L\dot\alpha}, Q^+_{R\beta}\equiv Q_{R\dot\beta}$. The generators $Q^+_{L\dot\alpha}, Q^+_{R\dot\beta}$ form a basis of complex-conjugate antichiral (antiholomorphic) $N=(\frac{1}{2}, \frac{1}{2})$ superalgebra obtained by Hermitean conjugation of relations (\ref{a8}-\ref{a9}). One can introduce the following outer antilinear antiautomorphism of fourth order (i.e. pseudoconjugation) of $N=(\frac{1}{2}, \frac{1}{2})$ \cite{c21,c22}
\begin{equation}\label{d3}
     \tau(Q_{L\alpha})=\epsilon_{\alpha\beta}Q^+_{L\beta}\qquad  \tau(Q_{R\alpha})=\epsilon_{\alpha\beta}Q^+_{R\beta}
\end{equation}
which maps the chiral $N=(\frac{1}{2}, \frac{1}{2})$ Euclidean superalgebra into antichiral one \footnote{The pseudoconjugation property $\tau^2=-1$ follows if $(Q^*)^+=(Q^+)^*$ what implies that the antichiral supercharges  $Q^+_{L\alpha}$ ,$Q^+_{R\beta }$ are mapped into $Q_{L\alpha}$ ,$Q_{R\beta }$ by the formula with the same algebraic form (\ref{d3}).}. If we use the relation (\ref{d2}) one can show easily that both superalgebras will have the same algebraic form (\ref{a8}-\ref{a9}) if the fourmomenta generators $\mathcal{P}_\mu$ are Hermitean (real).
The pseudoconjugation (\ref{d3}) has the form of quaternionic pseudoconjugation (\ref{a7b}), 
applied to two-component complex spinorial supercharges.

\subsection{Twisted  quantum deformations od $D=4$ $N=(\frac{1}{2},\frac{1}{2%
})$ holomorphic Euclidean superalgebra.}

In this subsection we shall consider the Euclidean superalgebra (\ref{a8}-\ref{a9}), invariant under the
pseudoconjugation (\ref{d1}) what implies that the fourmomenta $\mathcal{P}^{\mu }$  are real and Euclidean.

We shall study further the superextensions of the classical $D=4$ Euclidean $%
r$-matrices $r_{A}$ $\ (A=1,..,8)$ listed in formulae (\ref{a5})%
\begin{equation}\label{a11}
r_{A}\rightarrow r_{A}^{SUSY}=r_{A}+s_{A}
\end{equation}
where $s_{A}$ is the odd part of the supersymmetric $r$-matrix which is
bilinear in $Q_{L\alpha}$ and $Q_{R\beta }$. The general complex formula
for $s_{A}$ is the following (we
define $Q\vee Q^{\prime }=Q\otimes Q^{\prime }+Q^{\prime }\otimes Q)$
\begin{equation}\label{a12}
s_{A}=s_{A}^{(1)\alpha \beta}Q_{L\alpha}\vee Q_{L\beta}+s_{A
}^{(2)\alpha\beta}Q_{L\alpha}\vee Q_{R\beta }+s_{A}
^{(3)\alpha\beta}Q_{R\alpha }\vee Q_{R\beta }
\end{equation}
If we substitute (11) into the graded classical \ $YB$ equation and use the $%
D=4$ Euclidean superalgebra relations $(M_{i}^{\pm }=\frac{1}{2}(M_{i}\mp
N_{i})$
\begin{eqnarray}
\lbrack M_{i}^{+},Q_{L\alpha}]&=&-\frac{1}{2}(\sigma _{i})_{\alpha }^{\beta
}Q_{L\beta}\qquad \lbrack M_{i}^{-},Q_{L\alpha}]=[M_{i}^{+},Q_{R\alpha
}]=0\nonumber\\
\lbrack M_{i}^{-},Q_{R\alpha }]&=&\frac{1}{2}(\sigma _{i})_{\alpha }^{\beta
}Q_{R\beta ;}\label{a13}
\end{eqnarray}
one gets that the classical $r-$matrices $r_{A}$ (see (\ref{a5})) for $A=1,...,5$
do have a common superextension $s\equiv s_{A}$ given by the formula
\begin{equation}\label{a14}
 s=\eta Q_{L 2}\vee Q_{L 1}
\end{equation}
and for $A=6,7,8$ we obtain another three-parameter superextension
\begin{equation}\label{a15}
 \widetilde{s}=\eta ^{\alpha \beta }Q_{L\alpha}\vee Q_{L\beta }
\end{equation}
It should be added that the formulae (\ref{a14}-\ref{a15}) after the replacement $Q_{L\alpha}\rightarrow Q_{R\alpha}$
provide right-handed chiral extensions of $r-$matrices (\ref{a5}), which  as well describe the supersymmetrization
of (\ref{a11}) satisfying the supersymmetric YB equation.

The quantum deformations described by classical $r-$matrices $r_{6},$ $r_{7},
$ $r_{8}$ describe the  canonical twist deformations, which leads to $\mathbb{C}
-$number commutator of quantum space-time coordinates and  deforms Grassmann
superspace variables into the generators of $o(4)$ Clifford algebra \cite{c7} .
More complicated are the  the deformations described by $r_{1},...,r_{5}$.
The first, denoted  by $r_{1}$, generates the quadratic deformation of
Euclidean space-time algebra and the remaining four $r_{2},$ $r_{3},$ $%
r_{4,}r_{5}$ lead to Lie-algebraic formulae for non-commutative Euclidean
space-time algebra. In next paragraph we shall describe in detail the
deformation described by superymmetric classical $r$-matrix $r_{2}^{SUSY}$
(see (\ref{a5}), (\ref{a11}) and (\ref{a14})).

\section{$D=4$ Euclidean Hopf algebra providing Lie-algebraic deformation of
the bosonic Euclidean space sector.}\label{ab:sec3}

We shall consider now in detail quantum twisted Euclidean supersymmetry
generated by the classical $r-$matrix $r_{2}$ (see (\ref{a5})), and
supersymmetrically extended by the term (\ref{a14}). The algebraic sector of
Euclidean twisted superalgebra remains undeformed (see (\ref{a3}) but the coproducts are
modified by the similarity transformation \cite{Drinfeld}
\begin{equation}\label{f1}
\Delta =F^{-1}\Delta _{0}F
\end{equation}
where $\Delta _{0}$ describes the undeformed primitive coproduct. The twist
factor $F\in U(\mathit{E}_{4})\otimes U(\mathit{E}_{4})$ corresponding to $%
r_{2}$ looks as follows \footnote{In this section we shall drop the index $L$, i.e. $Q_{L\alpha}\rightarrow Q_{\alpha}$.} .
Here $\xi, \zeta_1, \zeta_2$ and $\eta$ are real deformation parameters in order to assure invariance ("unitarity") under the Euclidean Hermitean conjugation in the bosonic sector and pseudoconjugation (\ref{d1}) in the fermionic one. We get
\begin{equation}\label{f2}
F=\exp (i\xi M_{3}\wedge P_{0})\exp (i\zeta _{1}P_{3}\wedge P_{0}+i\zeta
_{2}P_{2}\wedge P_{1}+\eta Q_{1}\vee Q_{2})
\end{equation}\label{f3}
The  fourmomentum coproducts are the following
\begin{eqnarray}
\Delta (P_{0})&=&\Delta _{0}(P_{0})=\mathbf{1}\otimes
P_{0}+P_{0}\otimes \mathbf{1} \nonumber\\
\Delta (P_{j})&=&P_{j}\vee \cos (\xi P_{0})+\sin (\xi P_{0})\wedge
\epsilon _{3jk}P_{k},\qquad j,k=1,2,3
\end{eqnarray}
The coproducts for the "space" rotation generators are 
\begin{eqnarray}
\Delta (M_{j}) &=&M_{j}\vee \cos (\xi P_{0})+\sin (\xi P_{0})\wedge
\epsilon _{3jk}M_{k} +\delta _{j3}\{\Delta _{0}(M_{3})-M_{3}\vee \cos (\xi P_{0})\} \nonumber \\
&-&\zeta _{2}\{\epsilon _{j1k}P_{k}\cos (\xi P_{0})\wedge P_{2}\cos
(\xi P_{0})-\epsilon _{j1k}P_{k}\sin (\xi P_{0})\vee P_{1}\cos (\xi
P_{0})  \nonumber \\
&+&P_{1}\cos (\xi P_{0})\wedge \epsilon _{j2k}P_{k}\cos (\xi
P_{0})-P_{2}\cos (\xi P_{0})\vee \epsilon _{j2k}P_{k}\sin (\xi P_{0})\}
\nonumber \\
&-&\zeta _{1}\{\epsilon _{j3k}P_{k}\vee P_{0}\cos (\xi P_{0})+P_{0}\sin
(\xi P_{0})\wedge P_{j}\}+\frac{\eta}{2}R_j \label{f4}
\end{eqnarray}%
where $R_j$ is a shortcut for the following supersymmetric contribution 
\begin{eqnarray*}%
&R_j& =(\sigma _{j})_2^{\beta }\{Q_{\beta }\cos (%
\frac{\xi }{2}P_{0})\vee Q_{1}\cos (\frac{\xi }{2}P_{0})+
iQ_{\beta }\sin (\frac{\xi }{2}P_{0})\wedge Q_{1}\cos(%
\frac{\xi }{2}P_{0})\} \nonumber\\
&+&(\delta_{j3}\delta_2^\beta-i\epsilon_{j3k}(\sigma _{k})^\beta_2) \{iQ_{\beta }\cos(%
\frac{\xi }{2}P_{0})\wedge Q_{1}\sin(\frac{\xi }{2}P_{0})+\\
&+& Q_{\beta }\sin (\frac{\xi }{%
2}P_{0})\vee Q_{1}\sin (\frac{\xi }{2}P_{0}) \}
+(\sigma_{j})_1^{\beta }\{Q_{2}\cos(\frac{\xi }{2}P_{0})\vee Q_{\beta
}\cos(\frac{\xi }{2}P_{0})+\\
&+&iQ_{2}\cos(\frac{\xi }{2}P_{0})\wedge Q_\beta
\sin(\frac{\xi}{2}P_{0})\}
-(\delta_{j3}\delta_1^\beta-i\epsilon_{j3k}(\sigma _{k})^\beta_1)\times\\
&\times&\{i Q_{2}\sin(\frac{\xi}{2}P_{0})\wedge
Q_{\beta}\cos(\frac{\xi }{2}P_{0})
+Q_{2}\sin(\frac{\xi }{2}P_{0})\vee Q_{\beta }\sin(\frac{\xi }{2}P_{0})\}.\nonumber
\end{eqnarray*}%
The coproducts for Euclidean "boosts" are the following $(j=1,2,3)$
\begin{eqnarray}
&\Delta (N_{j}) &=N_{j}\vee \cos (\xi P_{0})+\sin (\xi P_{0})\wedge
\epsilon _{3jl}N_{l}-\xi \{\cos (\xi P_{0})\wedge P_{j}\nonumber\\
&+&\sin(\xi P_{0})\wedge \epsilon _{3jk}P_{k}\}
+\zeta _{2}\delta _{2j}\{P_{0}\cos (\xi P_{0})\vee P_{1}+P_{0}\sin
(\xi P_{0})\wedge P_{2}\}\nonumber\\
&-&\zeta _{1}\{\epsilon _{3jk}P_{k}\cos (\xi P_{0})\wedge P_{3}\sin (\xi P_{0})
 +P_{0}\cos (\xi P_{0})P_{3}\wedge P_{j}\cos (\xi P_{0})\}  \nonumber\\
&-&\zeta _{2}\delta _{1j}\{P_{0}\cos (\xi P_{0})\vee P_{2}-P_{0}\sin
(\xi P_{0})\wedge P_{1}\}+\frac{\eta}{2}R_j\label{f5}
\end{eqnarray}%
One can observe that for the generators $M_j^\pm$ (see (\ref{a13})) the $R_j$ (supersymmetric) contribution appears only in the antichiral sector $M^-_j$ while the chiral one remains bosonic. In more explicit form  one, e.g., gets 
\begin{eqnarray*}%
&R_1& = Q_{1 }\cos (\frac{\xi }{2}P_{0})\vee Q_{1}\cos (\frac{\xi }{2}P_{0})+
Q_{2}\cos (\frac{\xi }{2}P_{0})\vee Q_{2}\cos (\frac{\xi }{2}P_{0})\nonumber\\
&-&Q_{1 }\sin (\frac{\xi }{2}P_{0})\vee Q_{1}\sin (\frac{\xi }{2}P_{0})
-Q_{2 }\sin (\frac{\xi }{2}P_{0})\vee Q_{2}\sin (\frac{\xi }{2}P_{0})\nonumber \\
&+&2iQ_{1 }\sin (\frac{\xi }{2}P_{0})\wedge Q_{1}\cos (\frac{\xi }{2}P_{0})
-2iQ_{2 }\sin (\frac{\xi }{2}P_{0})\wedge Q_{2}\cos (\frac{\xi }{2}P_{0}) \nonumber\\
&R_3& = iQ_{2 }\cos (\frac{\xi }{2}P_{0})\wedge Q_{1}\sin (\frac{\xi }{2}P_{0})+
iQ_{1}\cos (\frac{\xi }{2}P_{0})\wedge Q_{2}\sin (\frac{\xi }{2}P_{0})\nonumber\\
&-&iQ_{1 }\sin (\frac{\xi }{2}P_{0})\wedge Q_{2}\cos (\frac{\xi }{2}P_{0})
-iQ_{2 }\sin (\frac{\xi }{2}P_{0})\wedge Q_{1}\cos (\frac{\xi }{2}P_{0})\nonumber
\end{eqnarray*}%
For the supercharges we obtain ($Q_{R\alpha}\rightarrow \tilde Q_\alpha$)
\begin{eqnarray}
\Delta (Q_{\alpha})&=&Q_{\alpha}\vee \cos(\frac{\xi }{2}P_{0})+i(\sigma
_{3})_\alpha^ \beta Q_{\beta}\wedge \sin(\frac{\xi }{2}P_{0}) \label{f6}\\
\Delta (\tilde Q_{\alpha }) &=&\cos(\frac{\xi }{2}P_{0})\vee\tilde Q_{\alpha }+i\sin(\frac{%
\xi }{2}P_{0})\wedge(\sigma _{3})_\alpha ^\beta\tilde Q_{\beta }+ 4\eta S_{\alpha (12)}\label{f7}\end{eqnarray}
where $S_{\alpha (12)}=\frac{1}{2}\left(S_{\alpha 12}+S_{\alpha 21}\right)$ denotes the symmetrization and
\begin{eqnarray*}
&S_{\alpha 12}&= Q_{1}\wedge \cos(\frac{%
\xi }{2}P_{0})\{(\sigma _{0})_{2 \alpha }P_{0}+(\sigma _{3})_{2
\alpha }P_{3}\}\\
&+&i(\sigma _{3})_{1}^{ \beta }Q_{\beta}\vee \sin(\frac{\xi
}{2}P_{0})\{(\sigma _{0})_{2 \alpha }P_{0}+(\sigma _{3})_{2 \alpha
}P_{3}\} \\
&+&Q_{1}\cos (\xi P_{0})\wedge \cos(\frac{\xi }{2}%
P_{0})\{\sum_{i=1,2}(\sigma _{i})_{2 \alpha }P_{i}\}\\
&+&i(\sigma _{3})_{1}^
\beta Q_{\beta}\cos (\xi P_{0})\vee \sin(\frac{\xi }{2}%
P_{0})\{\sum_{i=1,2}(\sigma _{i})_{2 \alpha }P_{i}\} \\
&+&Q_{1}\sin (\xi P_{0})\vee \cos(\frac{\xi }{2}%
P_{0})\{\sum_{i=1,2}(\sigma _{i})_{2 \alpha}\epsilon
_{3ik}P_{k}\}\\
&+&i(\sigma _{3})_1 ^\beta Q_{\beta}\sin (\xi
P_{0})\wedge \sin(\frac{\xi }{2}P_{0})(\sum_{i=1,2}(\sigma _{i})_{2
\alpha}\epsilon _{3ik}P_{k})
\end{eqnarray*}
In particular, calculates
\begin{eqnarray*}
&S_{1\, 21}&= iQ_{2}\wedge \exp(-\frac{i\xi }{2}P_{0})\{ P_{0}-iP_{3}\}\\
&S_{2\, 12}&= iQ_{1}\wedge \exp(\frac{i\xi }{2}P_{0})\{ P_{0}+iP_{3}\}\\
\end{eqnarray*}
Now due to  unitarity of the twist (\ref{f2}) one has the following reality property
\begin{eqnarray*} \Delta(x^*) &=&\Delta(x)^*
\end{eqnarray*}
for the coproduct, where $x^*=x$ in the bosonic sector $(M_{\mu\nu}, P_\rho)$ and $Q^*_\alpha=\tau(Q_\alpha),\ \tilde Q^*_\alpha=\tau(\tilde Q_\alpha)$ (see (\ref{d1})) for the fermionic one. Indeed, it can be checked that $F^*_j=F_j$ and $S_{1\,21}^*=S_{2\,12}$, while $S_{2\,12}^*=-S_{1\,21}$.
Similarly,  $S_{1\,12}^*=S_{2\,21}$ ($S_{2\,21}^*=-S_{1\,12}$).

In what follows we shall use  Mandelstam realization (see Ref. \refcite{c7}) of undeformed $N=(\frac{1}{2},%
\frac{1}{2})$ Euclidean superalgebra on left chiral Euclidean superspace $(x^\mu, \theta^\alpha)$, where $x^\mu$ are real Euclidean space-time coordinates and $\theta_{L\alpha}\equiv \theta_{\alpha}$ describes (left) chiral Grassmannian spinors transforming under $SU_L(2)$ ($\partial _{\mu }\equiv \frac{\partial
}{\partial x^{\mu }},$ $\partial _{\alpha}\equiv \frac{\partial }{\partial
\theta ^{\alpha}})$%
\begin{eqnarray}
P_{\mu } &=&-i\partial _{\mu },\qquad M_{\mu \nu }=ix_{[\mu }\partial _{\nu ]}+%
\frac{1}{2}(\theta \sigma _{\mu \nu })^\alpha\partial _{\alpha} \label{f8}\\
Q_{\alpha} &=&i\partial _{\alpha},\qquad \tilde Q_{\beta}=2\theta ^{\alpha
}(\sigma^{\mu })_{\alpha\beta }\partial_{\mu }\label{f9}
\end{eqnarray}%
Further we insert (\ref{f1}) into the $\star-$product formula  of chiral Euclidean fields $%
\Phi (x_{\mu },\theta _{\alpha})$
\begin{equation}
\Phi (x_{\mu },\theta _{\alpha})\star \Psi (x_{\mu },\theta _{\alpha
}):=m[F^{-1}\rhd \Phi (x_{\mu },\theta _{\alpha})\otimes \Psi (x_{\mu
},\theta _{\alpha})]
\end{equation}
If  $\rhd $ describes in (24) the realization (23) of Euclidean superalgebra generators occurring in
$F^{-1}$ one obtains the formulae for deformed superspace relations
\begin{eqnarray}
\lbrack x^{\mu }, x^{\nu }]_{\star } &=&2i(\zeta_1\delta_3^{[\mu}\delta_0^{\nu]}+\zeta_2\delta_2^{[\mu}\delta_0^{\nu]})+
2i\xi(\delta_2^{[\mu }\delta_0^{\nu ]}x^{1}+\delta_0 ^{[\mu }\delta_1 ^{\nu ]}x^{2})  \label{f10}\\
\lbrack x^{\mu }, \theta^{\alpha}]_{\star } &=&
\xi \delta_0 ^{\mu } (\sigma_3)_{\beta}^{\alpha}\theta^\beta\label{f11}\\
\{\theta ^{\alpha }, \theta ^{\beta }\}_{\star } &=&-2\eta 
\delta_1 ^{(\alpha }\delta_2 ^{\beta )}\label{f12}
\end{eqnarray}%
We observe that all deformation parameters are real and dimensionfull. 
The first part of relation (\ref{f10}) obtained by putting $\xi=0$, together with the relation (\ref{f12}), provides canonical type of deformation, while the second part of (\ref{f10}) together with
(\ref{f11}) ($\xi\neq 0$) leads to space-time deformation of Lie-algebraic type.
We add that the relations (\ref{f11}-\ref{f12}) are invariant under conjugation if we extend the
pseudoconjugation (\ref{d1}) to the Grassmannian coordinates
$\theta^\alpha$, provided that the Euclidean spacetime coordinates $x^\mu$ are Hermitian.

We recall that first deformation of chiral $N=(\frac{1}{2}, \frac{1}{2})$ Euclidean supersymmetries was
proposed by Seiberg \cite{c23} with primary deformation introduced a priori
by the canonical modification (\ref{f12}) of the anti-commutativity of Grassmann variables $%
\theta ^{\alpha}$. In such a framework the quantum-deformed
supersymmetries do not appear, but the anticommutators of supercharges are
modified and it appears that half of the considered Euclidean supersymmetries are broken explicitly
\footnote{Such explicit breaking has been also called $N=\frac{1}{2}$ breaking \cite{c23}.}.

\section{Outlook}
In this lecture we presented mainly the results of our paper Ref. \refcite{c7} but
 we also provided new results in Section 3, where the
explicit coproducts for  twisted $D=4$, $N=(\frac{1}{2}, \frac{1}{2})$ Euclidean chiral Hopf
superalgebra are written down. This Hopf superalgebra as well as the corresponding superspace do satisfy the Euclidean reality condition provided the fermionic sector undergoes the pseudoconjugation  (\ref{d1}).
It shows that the case of $N=(\frac{1}{2}, \frac{1}{2})$ Euclidean SUSY corresponds to $N=1$
Poincar\'{e} supersymmetry, but as follows from Sect. 2.2, it can be also obtained from Poincar\'{e} case after doubling of supercharges (in $N=1$ Poincar\'{e} superalgebra  we have four real supercharges, and in $N=(\frac{1}{2}, \frac{1}{2})$ Euclidean  case four complex). This passage from Poincar\'{e} to Euclidean SUSY is consistent with
Osterwalder-Schrader analytic continuation procedure \cite{c24} from Poincar\'{e} invariant theories in Minkowski space into Euclidean invariant
theories in  Euclidean $D=4$ space, where the doubling of fermions is emphasized. It should be stressed that imposing the pseudoconjugation  (\ref{d1}) or (\ref{d3}) cannot be used for the reduction of degrees of freedom of $N=(\frac{1}{2}, \frac{1}{2})$ supercharges.

The similarity between Euclidean and Poincar\'{e} case can be achieved however in the cases of $N=(1, 1)$ Euclidean and $N=2$ Poincar\'{e} supersymmetries, in both cases described by four complex or eight real supercharges. This property follows from the introduction for $N=(1, 1)$ Euclidean SUSY the conjugation $\#$ ($\#^2=1$), which can be used for imposing the reality constraints. These reality constraints are called $SU(2)-$Majorana condition \cite{c14} and are the consequences of quaternionic structure of $D=4$ Euclidean space, as we demonstrate below.

In order to introduce the reality condition consistent with
the quaternionic structure one has to consider a pair of quaternions $q^{a}
$ ($a=1,2)$  and define the quaternionic conjugation $(q^a)^\#$ (($(q^a)^\#)^\#=q^a$) in terms of quaterionic pseudoconjugation
(\ref{a7b}) as follows
\begin{equation}\label{a7c}
(q^{a})^\#=-\epsilon^{ab}(q^{b})^H\qquad\Leftrightarrow \qquad(\mathbf{q}_{\alpha }^{a})^\#=\epsilon^{ab} \epsilon_{\alpha\beta}\bar{\mathbf{q}}^b_\beta
\end{equation}
The quaternionic reality condition  ($SU(2)$ Majorana reality condition \cite{c14}) takes the form
\begin{equation}\label{a7d}
q^a=(q^{a})^\# \qquad\Leftrightarrow \qquad\mathbf{q}_{\alpha }^{a} =\epsilon^{ab} \epsilon_{\alpha\beta}\bar{\mathbf{q}}^b_\beta
\end{equation}
If we wish to obtain real Euclidean $N=(1,1)$ superalgebra with real fourmomenta one should introduce
the pairs of left  and right chiral supercharges $(Q_{L\alpha}^{a}, Q_{R\alpha }^{a}),\ a=1,2
$ which satisfy the superalgebraic counterpart of the  reality condition (\ref{a7d}).
\begin{equation}\label{bb1}
    Q^a_{L\alpha}=\epsilon^{ab}\epsilon_{\alpha\beta}(Q^b_{L\beta})^\dagger\qquad
    Q^a_{R\alpha}=\epsilon^{ab}\epsilon_{\alpha\beta}(Q^b_{R\beta})^\dagger
\end{equation}

In our next paper we shall consider the real $N=(1,1)$ Euclidean superalgebra and its quantum twist deformations.
It should be added that by considering only the doublets of supercharges $Q^a_{L\alpha}$
 ($Q^a_{R\alpha}$) we obtain $N=(2, 0)$ chiral ($N=(0, 2)$ antichiral) Euclidean superalgebras which can be mapped into each other by the $N=2$ extensions of the pseudoconjugations (\ref{d3}).
\section*{Acnowledgements}
Three of the authors (A.B., J.L. and V.N.T. would like to thank Prof.
Cheng-Ming Bao and Prof. Mo-Lin Ge for warm hospitality at Nankai University
in Tianjin. Two of us (A.B. and J.L.) were supported by the Polish NCN Grant
2011/51/B/ST2/03354 and V.N.T  was partly supported by the RFBR  grant No.11-01-00980-a and the
grant No.12-09-0064 of the Academic Fund Program of the National Research
University Higher School of Economics.


\end{document}